# Mono- *versus* multi-phosphonic acid based PEGylated polymers for functionalization and stabilization of metal (Ce, Fe, Ti, Al) oxide nanoparticles in biological media

V. Baldim[a], A. Graillot[b], N. Bia[b], C. Loubat[b] and J.-F. Berret*[a]

[a]*Matière et Systèmes Complexes, UMR 7057 CNRS Université Denis Diderot Paris-VII, Bâtiment Condorcet, 10 rue Alice Domon et Léonie Duquet, 75205 Paris, France.*
[b]*Specific Polymers, ZAC Via Domitia, 150 Avenue des Cocardières, 34160 Castries, France.*

**Abstract**
For applications in nanomedicine, particles need to be functionalized to prevent protein corona formation and/or aggregation. Most advanced strategies take advantage of functional polymers and assembly techniques. Nowadays there is an urgent need for coatings that are tailored according to a broad range of surfaces and that can be produced on a large scale. Herein, we synthesize mono- and multi-phosphonic acid based poly(ethylene glycol) (PEG) polymers with the objective of producing efficient coats for metal oxide nanoparticles. Cerium, iron, titanium and aluminum oxide nanoparticles of different morphologies (spheres, platelets, nanoclusters) and sizes ranging from 7 to 40 nm are studied in physiological and in protein rich cell culture media. It is found that the particles coated with mono-functionalized polymers exhibit a mitigated stability over time (< 1 week), whereas the multi-functionalized copolymers provide resilient coatings and long-term stability (> months). With the latter, PEG densities in the range $0.2 - 0.5$ nm$^{-2}$ and layer thickness about 10 nm provide excellent performances. The study suggests that the proposed coating allows controlling nanomaterial interfacial properties in biological environments.

# 1. Introduction
Over the last decade engineered nanoparticles have been developed as therapeutic, diagnostic, and theranostic agents, leading to the development of nanomedicine.[1-5] Recent studies have shown however that nanomedicine has not met the initial expectations regarding translation to the clinics. In a literature survey, Wilhelm *et al.* have found that with regard to solid tumor targeting, around 99 % of engineered particles administered intravenously to rodents are cleared from the blood circulation and miss their targets.[6] Others reviews on translational nanomedicine were reported and similar conclusions were reached.[4,7] One reason brought forward to explain these results is related to the hurdles encountered to control nanomaterial interfaces with biological fluids, cells and tissues. Although many progresses have been made in nanoparticle functionalization, innovative solutions are still to be investigated.

An efficient way to coat particles makes use of polymers[1,3,5] and co-assembly methods, avoiding hence the difficulties posed by grafting-to techniques and surface-initiated living radical polymerization.[8,9] With co-assembly the chains adsorb spontaneously at the particle surfaces by single or multiple point attachments and form a diffuse layer.[10-15] This layer is of a few nanometers and represents a barrier against particle aggregation and protein adsorption.[16-19] The affinity toward the surface can be enhanced by the addition of specific chemical groups that can react with the surface. For metal and metal oxide nanoparticles, the most commonly used linkers are alcohol, acid, amine, silane and thiol compounds.[14,15,17,20-30] Within a few exceptions,[20,31-35] functional polymers are generally synthesized for a single type of particles and for a predetermined application. Nowadays, there is an urgent need for coating agents that can be tailored according to a wide range of surfaces and produced on a large scale. To our knowledge, such a multi-purpose and multi-substrate coating has not been yet examined in details.

Herein we explore the possibility to use of phosphonic acid groups as efficient linkers to different metal oxide nanocrystals. Previous reports on functionalization have shown on PEG polymers terminated with a mono or bi-phosphonic acid functional groups[36,37] can adsorb on particles of different composition and nature, such as calcium carbonate,[38] iron,[14,17,23,24,26-28,39,40] cerium,[41,42] and titanium[43,44] oxides. In the previous examples the functional polymers were obtained from different synthesis and had different structures. In the context of nanomedicine, special attention was paid to iron oxide that was assessed *in vivo* as imaging and therapeutic agents. Sandiford *et al.* exploited PEG conjugates containing a terminal bisphosphonic acid group to overcome the rapid sequestration of the MRI probes and to increase their circulation time *in vivo*.[13] More recently, our group proposed an alternative strategy to increase the number of polymer attachment points. This strategy resulted in the synthesis of copolymers where phosphonic acid groups and PEG chains are covalently grafted to a poly(methyl methacrylate) backbone. Proof of concept studies performed on cells and on small animals confirmed that this technology has the potential to improve the colloidal stability in biofluids,[14] prevent protein adsorption[40] and increase the circulation time *in vivo*.[39]

Here we aim to demonstrate that multi-phosphonic acid PEG copolymers are susceptible to coat and stabilize a broad range of metal oxide particles in protein rich culture media. The particles are made of cerium, titanium, iron and aluminum oxides, have different morphologies (spheres, platelets, nanoclusters) and their sizes are comprised between 7 and 40 nm. In parallel, we synthesize an ensemble of six phosphonic acid based PEG polymers; three of them carry a unique phosphonic acid group, while the remaining three are statistical copolymers with multiple anchors. It is found that the particles coated with mono-functionalized polymers exhibit a mitigated stability over time (< 1 week), whereas the multi-functionalized copolymers provide resilient coatings and long-term stability (> months).

## 2. Results and discussion
### 2.1. Polymer synthesis and characterization
In this work, an ensemble of six polymers was synthesized for the coating and functionalization of metal oxide nanoparticles. Three polymers are linear poly(ethylene glycol) chains of molecular weight 1000, 2000 and 5000 g mol$^{-1}$ terminated by a single phosphonic acid group, -PO(OH)$_2$. In the following, these mono-functionalized chains are abbreviated PEG$_{1K}$-Ph, PEG$_{2K}$-Ph and PEG$_{5K}$-Ph, respectively. The remaining three polymers are statistical copolymers obtained by free radical polymerization following a synthesis pathway described in previous publications.[14,25] Two of these copolymers consist in a poly(methyl methacrylate) backbone with multiple phosphonic acid groups and methyl terminated PEG lateral chains in the molar proportions (0.50:0.50). The PEGs have a molecular weight of 2000 and 5000 g mol$^{-1}$, leading to the acronyms MPEG$_{2K}$-MPh and MPEG$_{5K}$-MPh (here "M" refers to the methyl methacrylate end-group of each comonomer). The sixth polymer contains an equimolar amount of methyl and amine terminated PEG chains in addition to the phosphonic acid groups. The molar proportions of PEGs, amine modified PEGs and phosphonic acids are thus (0.25:0.25:0.50) for this terpolymer, later abbreviated as MPEG$_{2K}$-MPEGa$_{2K}$-MPh (where "a" refers to the amine PEG-terminal group). Details on the synthesis and $^1$H NMR characterization can be found in the M&M section and in **Supplementary Information S1**. The molecular structures of the polymers used are illustrated in Figure 1. The weight-averaged molecular weights $M_w^{Pol}$ are determined from static light scattering using Zimm plots.[45] For Zimm plots, the Rayleigh ratio $\mathcal{R}(c)$ of the dispersions is measured as a function of the polymer concentration and the intercept of $Kc/\mathcal{R}(c)$ versus $c$ gives the inverse molecular weight (**Supplementary Information S2**). Here $K$ is the scattering contrast determined from refractometry.[14,39] The copolymers MPEG$_{2K}$-MPh, MPEG$_{5K}$-MPh and MPEG$_{2K}$-MPEGa$_{2K}$-MPh were found to have molecular weights $M_w^{Pol}$ of 20300, 22000 and 29200 g mol$^{-1}$ respectively. Assuming a molar mass dispersity Đ = 1.8,[14] the number-averaged molecular weight $M_n^{Pol}$ was determined at 11300, 12200 and 16200 g mol$^{-1}$. From these values, the average number of phosphonic acid was estimated at 5.1, 2.3 and 6.7. These later results are summarized in Table 1.

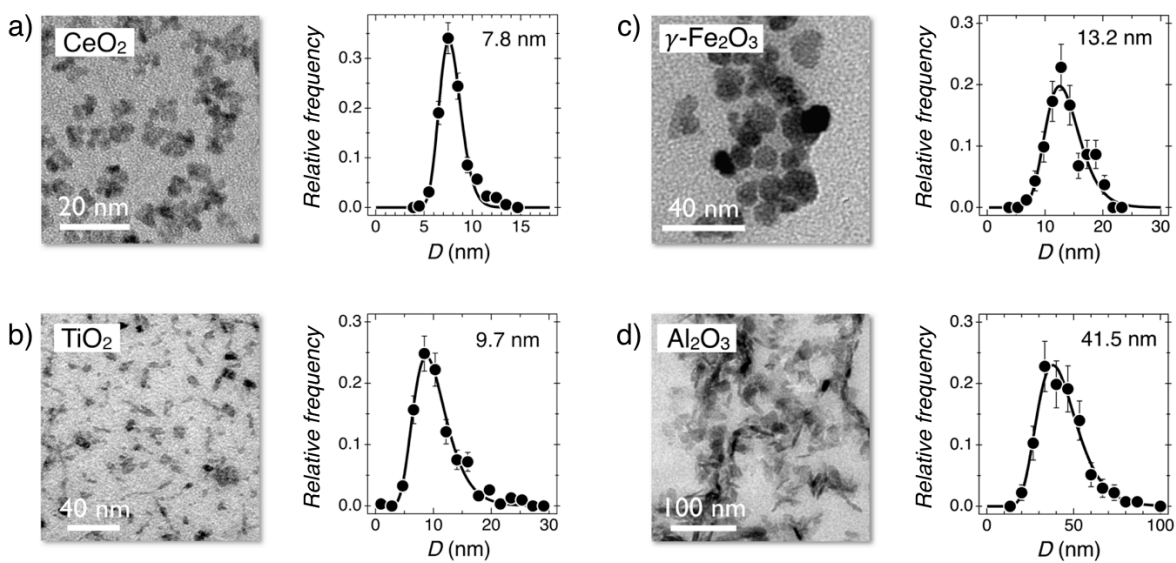

*Figure 1: Molecular structures of phosphonic acid based polymers and copolymers synthesized in this work. a) Poly(ethylene glycol) containing a terminal phosphonic acid group and PEG chains of 2000 g mol$^{-1}$ (PEG$_{2K}$-Ph). b) Poly(poly(ethylene glycol) methacrylate-co-dimethyl(methacryoyloxy)methyl phosphonic acid) is a statistical copolymer where the repeating units have lateral methyl terminated PEG$_{2K}$ chains and lateral phosphonic acids in the molar proportions (0.50:0.50) (MPEG$_{2K}$-MPh). c) MPEG$_{2K}$-MPEGa$_{2K}$-MPh is a statistical terpolymer where the repeating units have lateral methyl terminated PEG$_{2K}$ chains, lateral amine terminated PEG$_{2K}$ chains and lateral phosphonic acids in the molar proportions (0.25:0.25:0.50). Mono-functionalized PEGylated chains PEG$_{1K}$-Ph and PEG$_{5K}$-Ph, as well as the statistical copolymer MPEG$_{5K}$-MPh were also synthesized and studied as coats.*

| Polymers | $M_w^{Pol}$ (g mol⁻¹) | $M_n^{Pol}$ (g mol⁻¹) | Comonomer proportions | Phosphonic acids/polymer |
|---|---|---|---|---|
| PEG$_{1K}$-Ph | 1200 | 1200 | - | 1.0 |
| PEG$_{2K}$-Ph | 2000 | 2000 | - | 1.0 |
| PEG$_{5K}$-Ph | 5000 | 5000 | - | 1.0 |
| MPEG$_{2K}$-MPh | 20300 | 11300 | (0.50:0.50) | 5.1 |
| MPEG$_{5K}$-MPh | 22000 | 12200 | (0.50:0.50) | 2.3 |
| MPEG$_{2K}$-MPEGa$_{2K}$-MPh | 29200 | 16200 | (0.25:0.25:0.50) | 6.7 |

*Table 1: Molecular characteristics of the phosphonic acid PEGylated polymers and copolymers synthesized in this work*

## 2.2. Nanoparticles

The metal oxide nanoparticles investigated herein are nanocrystals of cerium ($CeO_2$), titanium ($TiO_2$), iron ($\gamma$-$Fe_2O_3$) and aluminum ($Al_2O_3$). Figure 2a–d show TEM micrographs of the particles together with their size distributions. Cerium oxide nanoparticles (nanoceria) were synthesized by thermo-hydrolysis of cerium nitrate salt under hydrothermal conditions as a 200 g L⁻¹ dispersion.[41] The particles consist in small agglomerates of 3 nm crystallites showing a median TEM size of 7.8 nm.[42,46,47] $TiO_2$ nanoparticles were obtained from Nano-structured & Amorphous Material Inc (Houston, TX, USA) as a 150 g L⁻¹ dispersion. The particles have the shape elongated platelets of median length 9.7 nm. Some small size aggregates are also observed in the TEM image.[48,49] Iron oxide nanoparticles (maghemite) were synthesized by alkaline co-precipitation of iron(II) and iron(III) salts and further oxidation.[46,50,51] The $\gamma$-$Fe_2O_3$ particles are superparamagnetic and display a slight shape anisotropy and a median size of 13.2 nm.[39] The $Al_2O_3$ nanoparticles are from Disperal® (SASOL, Hanmburg, Germany) and provided in the form of a white powder, that is dispersed by sonication and pH adjustment. Particles have the shape of irregular platelets of sizes 41.5 nm in length and 10 nm in thickness.[52,53] The $CeO_2$, $TiO_2$, $\gamma$-$Fe_2O_3$ and $Al_2O_3$ nanoparticles were also studied by dynamic light scattering, revealing hydrodynamic diameters $D_H$ of 9.6, 46.2, 29.8 and 55.3 nm, respectively. For cerium, iron and aluminum oxide nanoparticles, the differences found between the geometric and hydrodynamic diameters are attributed to the particle size and shape dispersity. For titanium oxide, it is ascribed to the presence of sub-100 nm aggregates in the dispersion. All dispersions were prepared in diluted nitric acid at pH 1.5, where these nanoparticles are positively charged and have zeta potentials $\zeta$ around + 30 mV.

## 2.3 – Polymer coated metal oxide nanoparticles

### 2.3.1 – Polymer coated cerium oxide nanoparticles

We first describe the polymer coating protocol focusing on $CeO_2$ and study the pH-stability diagrams. In a first step, polymer and particle stock solutions are prepared in the same conditions of pH and concentration. The dispersions are then mixed at different ratios $X = c_{NP}/c_{Pol}$, where $c_{NP}$ and $c_{Pol}$ are the nanoparticle and polymer concentrations and $c = c_{NP} + c_{Pol}$ remains constant. In practice, $c_{NP}$ is held in the range 0.1 – 10 g L⁻¹ and $X$ between 10⁻³ and 10³. For cerium oxide, the pH is set at 1.5 for the mixing and later increased to pH 8. The total concentration is fixed at $c = 2$ g L⁻¹. Figure 3a-f show the pH-stability diagrams obtained by plotting the hydrodynamic diameter $D_H$ as a function of $X$. For the three PEG$_{nK}$-Ph polymers (with $n = 1$, 2 and 5), the stability diagrams exhibit similar features (Figure 3a-c). At pH 1.5, $D_H(X)$ decreases continuously from a value around 20 nm to the diameter of the bare particles, $D_H = 9.6$ nm. At pH 8 in contrast, the diagrams show two regions: on the left-hand side the particle size remains stable (the $D_H$'s are identical to those of pH 1.5), whereas on the right-hand side the particles aggregate (shaded area). The limit between the two domains defines the critical ratio $X_C$. For the copolymers MPEG$_{2K}$-MPh, MPEG$_{5K}$-MPh and MPEG$_{2K}$-MPEGa$_{2K}$-MPh (Figure 3d-f), we find that in the range $X = 1 - 10$, the nanoceria are subjected to a partial aggregation. This aggregation arises from polymer bridging, in which phosphonic acid groups coming from a single chain adsorb on different particles.[16] The critical ratios are found around 1, except for MPEG$_{2K}$-MPh and MPEG$_{2K}$-MPEGa$_{2K}$-MPh where it is 1.5 ± 0.2 and 0.6 ± 0.1 respectively (Table 2).

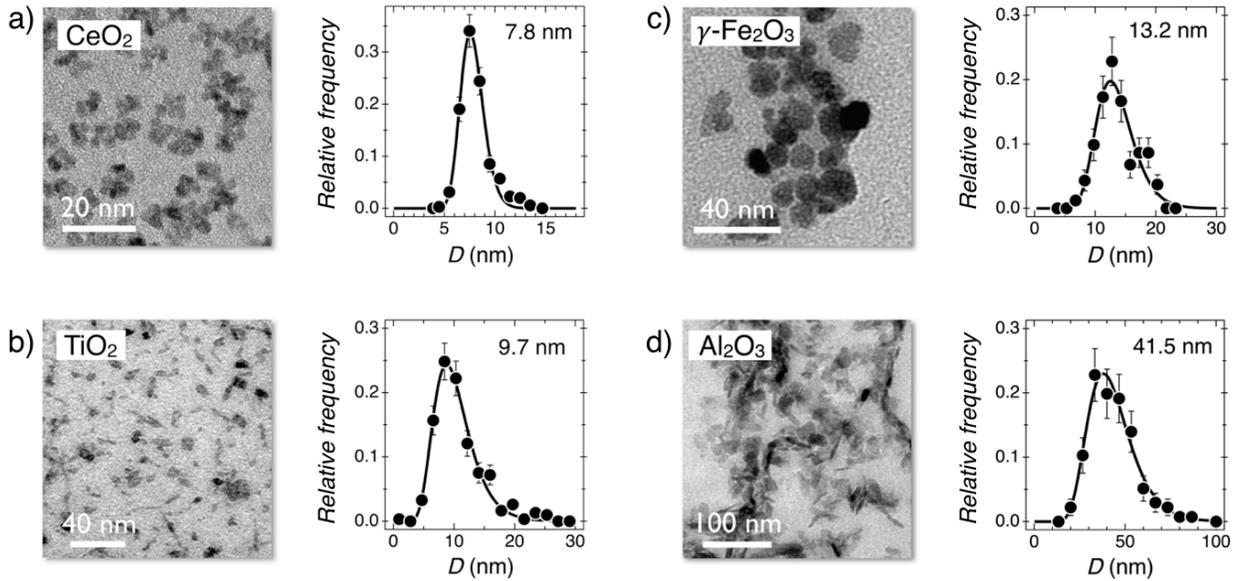

*Figure 2: Representative micrographs of metal oxide nanoparticles obtained by TEM and respective size distributions for **a)** CeO$_2$; **b)** TiO$_2$; **c)** γ-Fe$_2$O$_3$; **d)** Al$_2$O$_3$. The continuous curves are the results of best fit calculations using a log-normal function. The sizes indicated with the distributions represent the median particle diameters.*

The interpretation of the data of Figure 3 relies on the non-stoichiometric adsorption model developed by us in the context of polymer coating.[41] This model assumes that the polymers adsorb spontaneously on cerium oxide thanks to the phosphonic acid groups anchoring at the surface. This adsorption results in the stretching of the PEG chains, leading to the formation of a brush. The association is described as non-stoichiometric because the number of polymers adsorbed per particle depends on $X$. It is maximum below $X_C$ and decreases above. In case of partial coverage ($X > X_C$), the particles behave as uncoated CeO$_2$ and precipitate upon pH change.

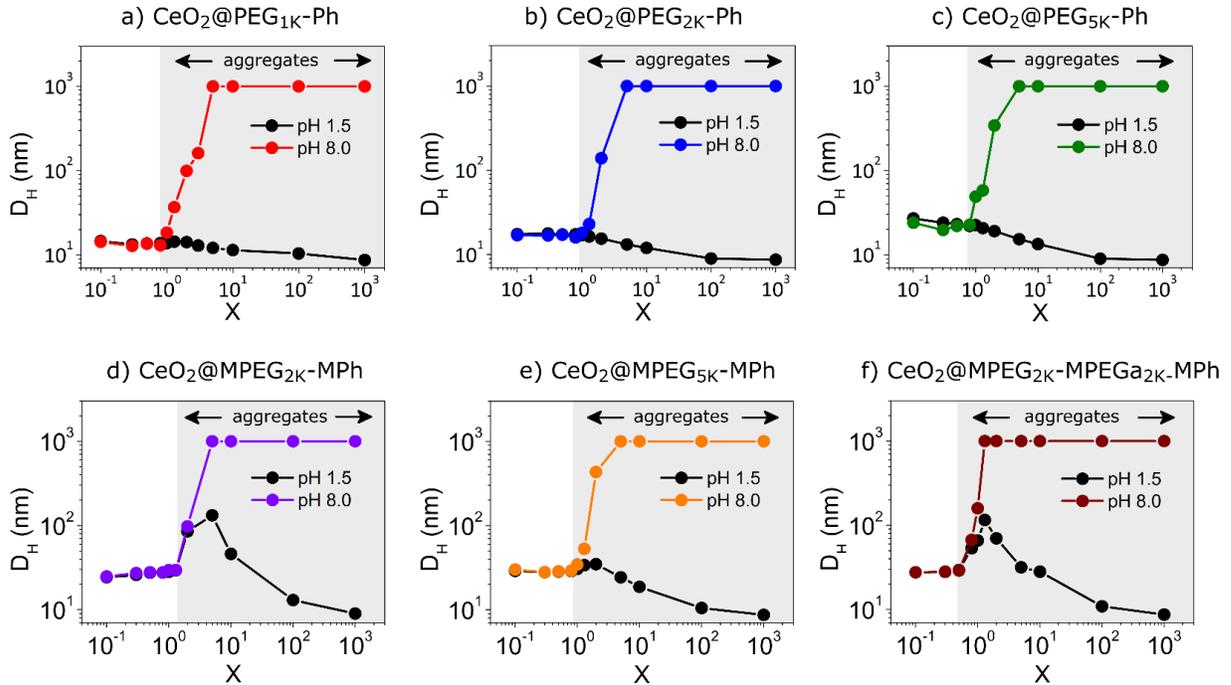

*Figure 3: Stability diagram of cerium oxide nanoparticles in presence of phosphonic acid PEG polymers (a-c) and copolymers (d-f) at pH 1.5 and 8. The hydrodynamic diameter $D_H$ measured by dynamic light scattering is shown as a function of the mixing ratio $X = c_{NP}/c_{Pol}$. The nanoceria are associated with: (a) $PEG_{1K}$-Ph, (b) $PEG_{2K}$-Ph, (c) $PEG_{5K}$-Ph, (d) $MPEG_{2K}$-MPh, (e) $MPEG_{5K}$-MPh, and (f) $MPEG_{2K}$-$MPEGa_{2K}$-MPh. For $X > X_C$, particles aggregate when pH is increased from 1.5 to 8. For micron-sized aggregates, $D_H$ is set at 1000 nm (shaded area). In the different figures, the bare nanoparticle solutions are set at $X = 10^3$ for convenience.*

According to the model, $X_C$ is linked to the number of polymers per particle $n_{Pol/NP}$ through the relationship $n_{Pol/NP} = (1/X_C)M_n^{NP}/M_n^{Pol}$, where $M_n^{NP}$ and $M_n^{Pol}$ are the number-averaged molecular weights of nanoceria and polymers, respectively. For the mono-functionalized polymers, the PEG density decreases from 0.85 to 0.38 and 0.20 nm$^{-2}$ with increasing molecular weight (Table 2). This result is consistent with the densities determined by quartz crystal microbalance with dissipation monitoring (QCM-D) on iron oxide flat substrates.[40] From QCM-D, it was concluded that the difference in PEG densities was arising from excluded volume effects and steric repulsion. During the film formation, the already adsorbed chains act as a barrier for the incoming ones, a mechanism that is more effective for longer chains. For the copolymers, the PEG densities are in the range 0.2 – 0.3 PEG nm$^{-2}$, in agreement with those of literature.[1,20,35,54,55] The value for $MPEG_{2K}$-$MPEGa_{2K}$-MPh is slightly larger (0.67 nm$^{-2}$), probably due to an underestimation of the critical ratio because of the bridging effects discussed previously. Thermogravimetric analysis (TGA) was performed on nanoceria powder samples coated with $MPEG_{2K}$-MPh and $MPEG_{2K}$-$MPEGa_{2K}$-MPh and revealed PEG densities of 0.12 and 0.18 nm$^{-2}$ respectively (**Supplementary Information S3**). These values are slightly lower that those derived from $X_C$, indicating that the stability diagram determination probably overestimates the amount of adsorbed polymers. The average number of polymers per $CeO_2$ and the PEG densities obtained are listed in Table 2.

| Particles | $X_C$ | Polymers per particle | PEG chain density (nm$^{-2}$) |
|---|---|---|---|
| CeO$_2$@ PEG$_{1K}$-Ph | 0.9 ± 0.1 | 230 | 0.85 |
| CeO$_2$@ PEG$_{2K}$-Ph | 1.2 ± 0.2 | 103 | 0.38 |
| CeO$_2$@ PEG$_{5K}$-Ph | 0.9 ± 0.1 | 55 | 0.20 |
| CeO$_2$@ MPEG$_{2K}$-MPh | 1.5 ± 0.2 | 15 | 0.28 |
| CeO$_2$@ MPEG$_{5K}$-MPh | 0.9 ± 0.1 | 23 | 0.20 |
| CeO$_2$@ MPEG$_{2K}$-MPEGa$_{2K}$-MPh | 0.6 ± 0.1 | 26 | 0.62 |

**Table 2**: *Critical mixing ratio $X_C$, average number of polymers per particle and PEG densities on polymer coated cerium oxide nanoparticles.*

Another approach for testing the non-stoichiometric model consists in examining the $X$-dependence of the scattered intensity. This intensity is transposed into the Rayleigh ratio $\mathcal{R}(X)$ normalized with that of the nanoceria dispersion $\mathcal{R}_{NP}$ at $c = 2$ g L$^{-1}$. Figure shows the quantity $\mathcal{R}(X)/\mathcal{R}_{NP}$ as a function of $X$ for the six polymers investigated. For the PEG$_{nK}$-Ph in Figure 4a, the normalized Rayleigh ratio first increases with increasing $X$, passes through a maximum around $X_C$ and then decreases to 1. The continuous lines are best fit calculations using the model equations,[41] as described in **Supplementary Information S4**. The agreement with the model is excellent for the 3 polymers. The principal fitting parameter is the number of polymers per particle $n_{Pol/NP}$ that determines the maximum position. The $n_{Pol/NP}$ retrieved from the fitting confirm those obtained from the $X_C$ determination in Table 2. With the copolymers, the model fails to account for the scattering intensity, as illustrated in Figure 4b. A good agreement is achieved at low and high $X$-values, but not in the region where the dispersions show aggregation. For these polymers, we rely on the densities determined from the $X_C$. In the upcoming section, coated CeO$_2$ particles are prepared with a large excess of polymers ($X = X_C/5$) and later dialyzed against DI-water to remove this excess. The dispersions were concentrated to 20 g L$^{-1}$ and stored in the fridge, where they display long-term stability (> months).

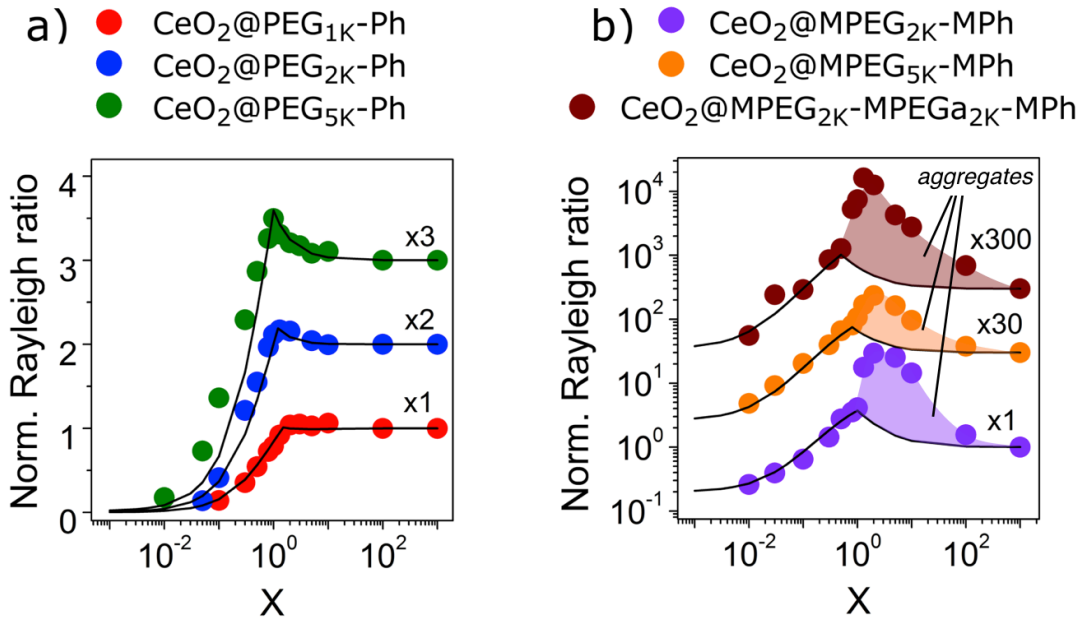

*Figure 4: Normalized Rayleigh intensity $\mathcal{R}(X)/\mathcal{R}_{NP}$ as a function of X for CeO$_2$ nanoparticles and phosphonic acid PEG polymers at pH 1.5:* **a)** *PEG$_{1K}$-Ph, PEG$_{2K}$-Ph and PEG$_{5K}$-Ph;* **b)** *MPEG$_{2K}$-MPh, MPEG$_{5K}$-MPh and MPEG$_{2K}$-MPEGa$_{2K}$-MPh. The concentration is set at 2 g L$^{-1}$. The continuous lines are best fit calculations using the non-stoichiometric interaction model developed in* **Supplementary Information S4**.[41]

### 2.3.2 Polymer coated maghemite, titania and alumina nanoparticles

The previous strategy was implemented with the titanium, iron and aluminum oxide particles. Table 3 shows the hydrodynamic diameter $D_H$ for the four polymer coated metal oxide nanoparticles, together with the brush thickness $h = \frac{1}{2}\left(D_H^{coated} - D_H^{bare}\right)$ derived for the twenty-two polymer-particle configurations tested, where $D_H^{bare}$ and $D_H^{coated}$ stand for the hydrodynamic diameter of the bare and coated particles. It is found that whatever the particle size, the layer thickness remains the same for a given polymer. Such an outcome suggests that the adsorption mechanism and the brush conformation are similar and independent on the substrate. A closer look at Table 3 reveals that for the mono-functionalized polymers, $h$ increases with the PEG molecular weight, from 2.7 nm for PEG$_{1K}$-Ph to 4.5 nm for PEG$_{2K}$-Ph and 8.7 nm for PEG$_{5K}$-Ph. These values are close to those found for films obtained through QCM-D on flat iron surfaces.[40] For the block copolymers with multi-phosphonic acids, the film thickness is in the range 9 – 11 nm and only slightly molecular weight dependent. Figure 5a and 5b depict the conformation of the phosphonic acid end-groups regarding the surface for the mono- and multi-functionalized polymers, respectively. The inset shows the MPEG$_{2K}$-MPh bound to a nanoparticle together with the PEG brush. It has been known that phosphonic acids strongly bind to the surface of metal oxides through condensation of their acidic hydroxyls P-OH with surface metal hydroxyls metal-OH and/or through the coordination of the phosphoryl oxygen to Lewis acid surface sites.[36,37,56] Mono-, bi- or tridentate anchoring modes have been proposed in the literature and are illustrated Figure 5b.

| | | Polymer coatings | | | | | |
|---|---|---|---|---|---|---|---|
| **Particles** | Bare | PEG$_{1K}$-Ph | PEG$_{2K}$-Ph | PEG$_{5K}$-Ph | MPEG$_{2K}$-MPh | MPEG$_{5K}$-MPh | MPEG$_{2K}$-MPEGa$_{2K}$-MPh |
| CeO$_2$@ | 9.6 | 14.8 | 19.0 | 27.0 | 27.2 | 31.7 | 31.5 |
| TiO$_2$@ | 46.2 | - | 57.0 | 65.8 | 62.9 | 68.8 | 65.6 |
| γ-Fe$_2$O$_3$@ | 29.8 | 35.2 | 40.0 | 48.8 | 49.5 | 55.2 | - |
| Al$_2$O$_3$@ | 55.3 | 59.1 | 61.0 | 70.0 | 70.5 | 79.5 | 81.8 |
| **shell thickness** | - | 2.7 ± 0.6 | 4.5 ± 1.0 | 8.7 ± 1.0 | 8.9 ± 1.0 | 11.8 ± 0.7 | 11.3 ± 1.8 |

*Table 3: Hydrodynamic diameters of polymer coated nanoparticles and respective polymer layer thicknesses obtained in HNO$_3$ pH 1.5*

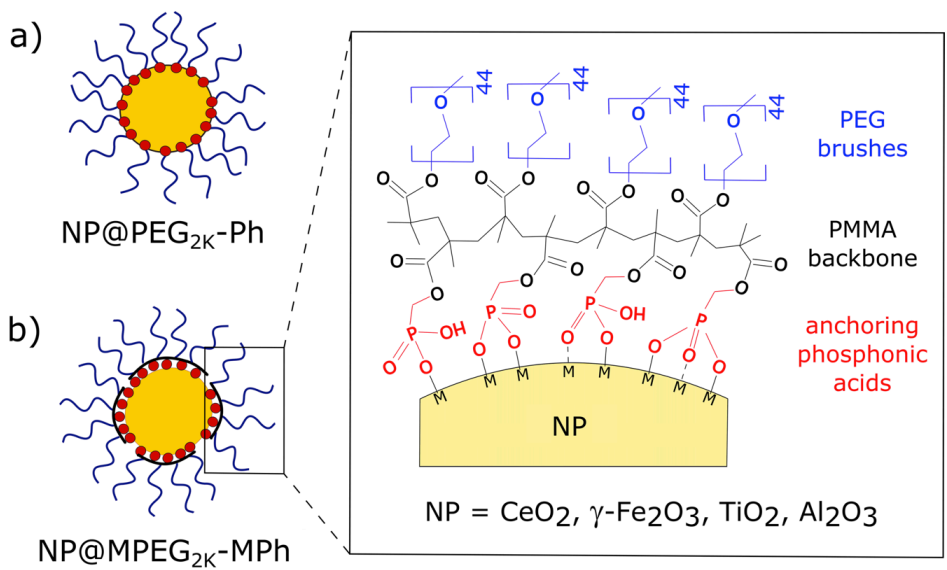

*Figure 5: a,b) Schematic representation of a nanoparticle coated with phosphonic acid based poly(ethylene glycol) polymers using mono- and multi-functionalized polymers respectively. Inset: Possible mono-, bi- or tridentate anchoring modes for phosphonic acid at metal oxide surfaces.*

## 2.4. Colloidal stability of bare and polymer coated oxide nanoparticles
### 2.4.1. Bare oxide nanoparticles

We now turn to the issue of the colloidal stability as a function of pH, ionic strength and protein content. Figure 6a and 6b display the pH-dependences of the hydrodynamic diameter $D_H$ and zeta potential $\zeta$ for the bare cerium, titanium, iron and aluminum oxide nanoparticles. At low pH, the dispersions are stable and the particle $D_H$'s are those of Table 3. The particles are stabilized by electrostatic repulsion mediated by the metal-$OH_2^+$ surface groups. As a result, the zeta potential is positive and around + 30 mV. With increasing pH, the charges present at the crystalline planes in contact with the solvent are gradually neutralized, leading to a decrease in density and electrostatic screening. The decrease of the zeta potential in Figure 6b is concomitant to the particle aggregation, which is revealed by the steep $D_H$-increase (Figure 6a). At physiological pH, the dispersions are turbid and precipitation is observed, in agreement with earlier studies.[50,57-61] The nanoparticle redispersion due to charge inversion at high pH is not observed, indicating the irreversible character of the aggregation process.

Figure 6c-e display images of the metal oxide dispersions in nitric acid at pH 1.5, in phosphate buffer (PBS) and in phenol red free Dulbecco's Modified Eagle's Medium (DMEM) complemented with 10% fetal bovine serum (FBS), respectively. At physiological pH, precipitates are observed at the bottom of the PBS and DMEM vials for the four dispersions. The precipitation is the result of a number of interacting factors. In PBS, it is due to pH and ionic strength effects related to the solvent change and to the modification of the inter-particle repulsion. In DMEM, proteins and other biological molecules also come into play and cause aggregation. Two movies in **Supplementary Information** illustrate the process kinetics and show that the phenomenon is rapid (< 1 s). There, 20 µL of a concentrated $CeO_2$ dispersion are added to PBS (**Movie#1**) and to protein enriched cell culture medium (**Movie#2**). As the drop reaches the solvent, particles aggregate immediately, as indicated by the growth of large scattering flakes within the actuated solution.

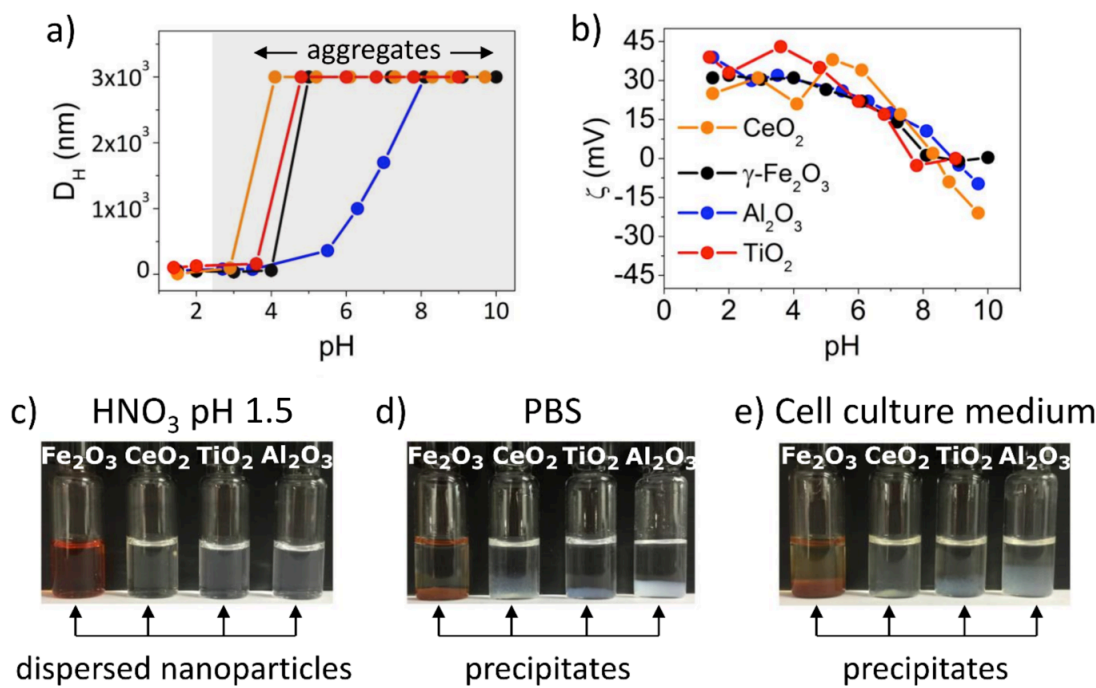

***Figure 6:*** *a) Hydrodynamic diameter $D_H$ and **b)** zeta potential $\zeta$ determined for the cerium, titanium, iron and aluminum oxide dispersions as a function of the pH. **c)** Images of vials containing γ-$Fe_2O_3$, $CeO_2$, $TiO_2$, $Al_2O_3$ dispersions in $HNO_3$, pH 1.5; **d)** same as in Figure 6c using phosphate buffer saline at pH 7.4; **e)** same as in Figure 6c using cell culture medium (here phenol red free Dulbecco's Modified Eagle's Medium complemented with 10% fetal bovine serum).*

### 2.4.2. Mono- *versus* multi-phosphonic acid PEG polymer coating

To assess the coating performances of the PEGylated polymers in physiological solvents, light scattering was performed to test the dispersion state over time. In the experiments performed, 20 μL of a 20 g L$^{-1}$ nanoparticle dispersion are diluted ten times in PBS or in complete cell culture medium. At the concentration of 2 g L$^{-1}$, it was verified that the scattering arising from the particles is much larger than that from the solvent, and that in the cell culture medium the protein contribution was negligible (see **Supplementary Information S5)**. The dispersions are then studied as a function of time at day 1, 2 and 7. A final assessment is realized at 60 days after mixing.

Figure 7a-c display the second-order autocorrelation function of the scattered light, $g^{(2)}(t)$ obtained in cell culture medium for $CeO_2$ coated with the mono-functionalized polymers. At day 1, the data exhibit a quasi-exponential decay associated with a unique relaxation mode. The hydrodynamic diameters are 14.8, 19.0 and 27.0 nm, corresponding to a polymer brush thickness of 2.6, 4.7 and 8.7 nm (Table 3). At day 2 and 7, $CeO_2$@PEG$_{1K}$-Ph and $CeO_2$@PEG$_{2K}$-Ph and $CeO_2$@PEG$_{5K}$-Ph show signs of aggregation, with autocorrelation function $g^{(2)}(t)$ shifting to the right hand-side and corresponding to the decrease of the diffusion constant. The intensity distributions in the insets also illustrate this augmentation. Note that the shorter the PEGs chains, the faster the kinetics. For the PEG$_{5K}$-Ph coat, the aggregation is seen only after two months (**Supplementary Information S6**). In contrast, nanoceria coated with the multi-functionalized copolymers have $g^{(2)}(t)$ that remain unchanged over time (Figure 7d-f). The autocorrelation functions exhibit again a unique relaxation mode at day 1, 2 and 7 associated hydrodynamic diameters of 27.2, 31.7 and 31.5 nm. For these samples, the brush

thickness is not altered by the presence of proteins and remains at the value found in DI-water, 8.9, 11.8 and 11.3 nm, respectively. These outcomes suggest that the coated nanoceria are devoid of a protein corona. Experiments performed using PBS under the same conditions show also a mitigated stability for the mono-functionalized polymer coatings and a resilient stability for the multi-functionalized copolymers.

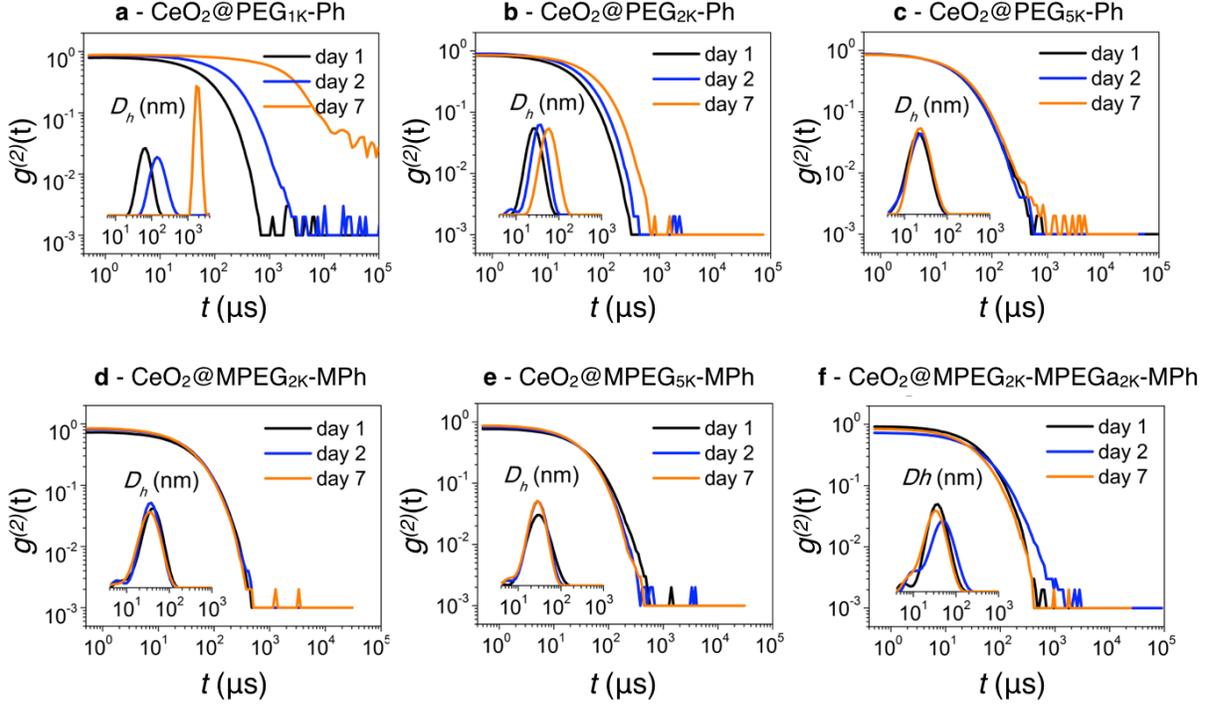

*Figure 7:* *Autocorrelation functions $g^{(2)}(t)$ obtained from dynamic light scattering on phosphonic acid PEG polymer coated $CeO_2$ nanoparticles as a function of the time. Experiments were performed in cell culture medium (Dulbecco's Modified Eagle's Medium, DMEM) complemented with 10% fetal bovine serum (FBS): a) $PEG_{1K}$-Ph; b) $PEG_{2K}$-Ph; c) $PEG_{5K}$-Ph; d) $MPEG_{2K}$-MPh; e) $MPEG_{5K}$-MPh; f) $MPEG_{2K}$-$MPEGa_{2K}$-MPh. The insets display the intensity distributions for the hydrodynamic diameter $D_H$. A shift of the $g^{(2)}(t)$ to the right-hand side of the diagram is an indicator of the protein corona formation or particle aggregation.*

The above results suggest that the colloidal behaviors observed in Figure 7 result from the interplay between different parameters. For the $PEG_{nK}$-Ph coating in PBS, the particle aggregation is attributed to the progressive removal of the PEGylated polymers from the surface. Due to this exchange, the effective ratio $X$ at the particle level increases and beyond the critical value $X_C$ the particles start to destabilize and aggregate (as in Figure 3). In cell medium, the displacement of the $PEG_{nK}$-Ph away from the surface also favors the protein adsorption, which again accelerates the destabilization kinetics. For $CeO_2$@$PEG_{1K}$-Ph it is also possible that the coating thickness (2.7 nm) is not sufficient to offset the attractive van der Waals forces.[62] In contrast, $CeO_2$ coated with the multi-phosphonic acid copolymers are stable both in PBS and complete DMEM for longer times. These results suggest that a multiple anchoring of phosphonic acid groups is more favorable for a resilient copolymer adsorption. We also show that the PEG functionalization by an amino group terminus does not modify this stability property. In conclusion, we have shown that uncoated nanoceria destabilize rapidly in physiological conditions (within less than 1 second), whereas multi-phosphonic acid PEG copolymers provide a resilient coating for months. In terms of layer structure, results show that PEG densities in the range 0.2 – 0.5 nm$^{-2}$ and PEG thickness about 10 nm provide excellent performances.

**2.4.3. Stabilizing titanium, iron and aluminum oxide dispersions in biological environments**

The previous stability assays were replicated with the titanium, iron and aluminum oxide dispersions and with the MPEG$_{2K}$-MPh copolymer. Concentrated dispersions of polymer coated metal oxide nanoparticles (20 g L$^{-1}$) were prepared at the ratio $X_C/5$ and pH 1.5, followed by a pH increase and ultracentrifugation. The $X_C$-values determined for each polymer-oxide pair are found at 1.5, 1.5, 1.5 and 3.5 for cerium, titanium, iron and aluminum nanoparticles. Figure 8a-d compare the time evolution of the autocorrelation functions $g^{(2)}(t)$ for the four metal oxide dispersions in the protein rich cell culture medium, CeO$_2$@MPEG$_{2K}$-MPh, TiO$_2$@MPEG$_{2K}$-MPh, γ-Fe$_2$O$_3$@MPEG$_{2K}$-MPh and Al$_2$O$_3$@MPEG$_{2K}$-MPh, respectively. The data for PBS can be found in **Supplementary Information S7**. In each panel, the insets display the corresponding intensity size distributions. Either in DMEM or in PBS, the data show an overall excellent colloidal stability and demonstrate that the phosphonic acid PEG copolymers are efficient coats for nanoparticle substrates of different sizes (7 – 40 nm) and chemical compositions. An excellent colloidal stability is also obtained with the amine containing terpolymer MPEG$_{2K}$-MPEGa$_{2K}$-MPh (**Supplementary Information S8**), paving the way to further functionalization, using for instance a targeting peptide or a contrast agent. Here again, the dispersion stability benefits from the multi-site attachment of the phosphonic acid groups at the particle surfaces.

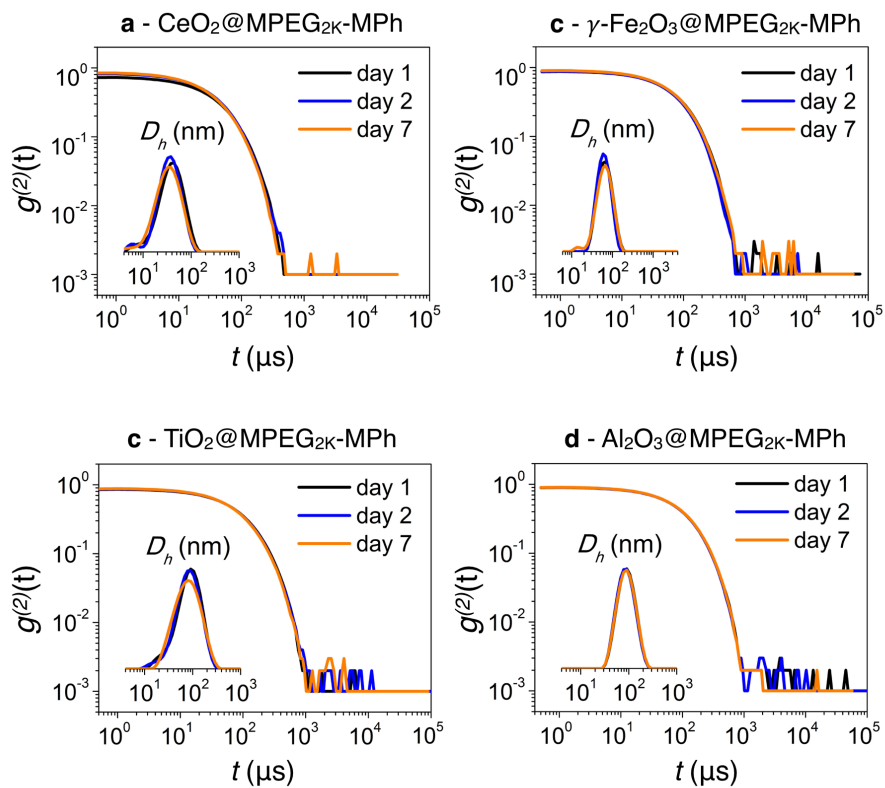

*Figure 8: Autocorrelation functions $g^{(2)}(t)$ obtained from dynamic light scattering on metal oxide nanoparticles coated with the copolymer MPEG$_{2K}$-MPh as a function of the time. Experiments were performed in cell culture medium (DMEM) complemented with 10% fetal bovine serum (FBS): a) CeO$_2$@MPEG$_{2K}$-MPh, b) TiO$_2$@MPEG$_{2K}$-MPh, c) γ-Fe$_2$O$_3$@MPEG$_{2K}$-MPh and d) Al$_2$O$_3$@MPEG$_{2K}$-MPh. Insets: Intensity distributions corresponding to the autocorrelation functions $g^{(2)}(t)$.*

# 4. Conclusion

Cerium, iron, titanium and aluminum oxide nanoparticles of different morphologies and sizes ranging from 7 to 40 nm are dispersed in a protein enriched cell culture medium and their colloidal stability is investigated by static and dynamic light scattering over time. These particles are coated with six polymers containing poly(ethylene glycol) chains and one or several phosphonic acids as anchoring moities, among them a terpolymer containing amine functionalized PEG chains suitable for further chemical modification. The strategy used for the preparation of polymer coated nanoparticles works well for all particles investigated and is based on mixing bare particle dispersions and polymer solution at acidic pH and mixing volume ratio smaller than a critical value $X_C$, allowing the production of large amounts of dispersions. It is found that the particles coated with mono-functionalized polymers exhibit a mitigated stability over time (< 1 week), whereas the multi-functionalized copolymers provide resilient coatings and long-term stability (> months). This multi-purpose and multi-substrate coating represents a step toward the understanding and control of nanomaterial interfacial phenomena with biological fluids, cells and tissues.

# 5. Materials and Methods

## 5.1. Materials

**Chemicals:** Phosphate buffer saline (PBS1X), trypsin–EDTA, DMEM and DMEM without phenol red (called phenol red free DMEM in the following), fetal bovine serum (FBS) and penicillin–streptomycin were purchased from Gibco, Life Technologies. The DMEM composition is shown in **Supplementary Information S9**. Water was deionized with a Millipore Milli-Q Water system. All products were used without purification.

**Nanoparticles:** Cerium oxide nanoparticles with a nominal diameter of 7.8 nm were synthesized and kindly given by Rhodia (Centre de Recherche d'Aubervilliers, Aubervilliers, France) as a 200 g $L^{-1}$ aqueous dispersion (pH 1.5).[47] Iron oxide nanoparticles were synthesized according to the Massart method by alkaline coprecipitation of iron(II) and iron(III) salts and oxidation of the magnetite ($Fe_3O_4$) into maghemite ($\gamma$-$Fe_2O_3$) giving a 20 g $L^{-1}$ aqueous dispersions at pH 2. Aluminum oxide nanoparticles (Disperal, SASOL) were kindly given by Florent Carn (Laboratoire Matière et Sytèmes Complexes, Paris). The powder was dissolved in $HNO_3$ pH 1.5 and sonicated for 30 min to give a 10 g $L^{-1}$ aqueous dispersion. Titanium oxide nanoparticles (anatase) with a nominal particle diameter of 15 nm were purchased as a 170 g $L^{-1}$ $TiO_2$ suspension in water from Nanostructured & Amorphous Material Inc. (Houston, TX, USA). The dispersion was provided by Serge Stoll from Geneva University.

**Polymers:** The mono- and multi-phosphonic acid PEG polymers were synthesized by Specific Polymers®, France (http://www.specificpolymers.fr/). Synthesis details can be found in previous reports.[14,39] For the copolymers, a molar-mass dispersity of 1.8 was obtained by size exclusion chromatography on PolyPore column using THF as eluent and polystyrene standards. The synthesis of the terpolymer $MPEG_{2K}$-$MPEGa_{2K}$-MPh is described in details in the **Supplementary Information S1.** It was characterized from $^1H$ NMR and $^{31}P$ NMR using a Bruker Advance 300 spectrometer operating at 300 MHz. From the molar equivalent of acid groups obtained from NMR and molecular weight determination, the number of phosphonic acids and PEG segments was estimated (Table 2). For $^1H$ NMR, chemical shifts were referenced to the peak of residual non-deuterated solvents at 7.26 ppm for $CDCl_3$, as detailed in **Supplementary Information S1**. The weight-averaged molecular weight $M_w$ of the PEG copolymers was determined by static light scattering measurements using a NanoZS

Zetasizer from Malvern Instrument. The polymer solutions were prepared with 18.2 MΩ Milli-Q water, filtered with 0.2 μm cellulose filters and their pH was adjusted to 8 by addition of ammonium hydroxide. The Rayleigh ratio was measured as a function of the concentration and the polymer molecular masses $M_w^{Pol}$ were determined through the Zimm representation, as detailed in **Supplementary information S2**. The $M_w^{Pol}$-values were 20300 g mol$^{-1}$, 22000 g mol$^{-1}$ and 29200 g mol$^{-1}$ for MPEG$_{2K}$-MPh, MPEG$_{5K}$-MPh and MPEG$_{2K}$-MPEGa$_{2K}$-MPh, respectively.

**Transmission electronic microscopy (TEM):** Micrographs were taken with a Tecnai 12 TEM operating at 80 kV equipped with a 1K×1K Keen View camera. Nanoparticle dispersions were deposited on ultrathin carbon type-A 400 mesh copper grids (Ted Pella, Inc.). Micrographs were analysed using ImageJ software for 200 particles. The particle size distributions are adjusted using a log-normal function of the form (Figure 2): $p(d, D, s) = \frac{1}{\sqrt{2\pi}\beta(s)d} exp\left(-\frac{ln^2(d/D)}{2\beta(s)^2}\right)$. In the previous equation, , $D$ is the median diameter and $\beta(s)$ is related to the size dispersity $s$ through the relationship $\beta(s) = \sqrt{ln(1+s^2)}$. $s$ is defined as the ratio between the standard deviation and the average diameter. For $\beta < 0.4$, one has $\beta \cong s$.[14,41]

**Static and dynamic light scattering:** Light scattering measurements were carried out using a NanoZS Zetasizer (Malvern Instruments) at detection angle at 173°. The hydrodynamic diameter $D_H$ and the zeta potential $\zeta$ were measured. The second-order autocorrelation function is analyzed using the cumulant and CONTIN algorithms to determine the average diffusion coefficient $D$ of the scatterers. Hydrodynamic diameter $D_H$ is then calculated according to the Stokes–Einstein relation, $D_H = k_B T/3\pi\eta_S D$ where $k_B$ is Boltzmann's constant, $T$ the temperature (298 K) and $\eta_S$ the solvent viscosity. Measurements were performed in triplicate at 25 °C after an equilibration time of 120 s. Viscosities of the solvents used can be found in Table 4.

| Solvent | Viscosity $\eta_S$ (mPa s) | Refractive index |
|---|---|---|
| H$_2$O | 0.8872 | 1.333 |
| Phosphate buffer saline (PBS1X) | 0.9103 | 1.332 |
| Cell culture medium (DMEM + 10% FBS) | 0.9400 | 1.345 |

*Table 4: Parameters used for static and dynamic light scattering measurements*

**Polymer coated nanoparticles:** Dispersions of CeO$_2$, γ-Fe$_2$O$_3$, TiO$_2$ and Al$_2$O$_3$ nanoparticles were diluted to concentration of 2 g L$^{-1}$ in HNO$_3$ (pH 1.5). Polymers solutions of PEG$_{1K}$-Ph, PEG$_{2K}$-Ph, PEG$_{5K}$-Ph, MPEG$_{2K}$-MPh, MPEG$_{5K}$-MPh and MPEG$_{2K}$-MPEGa$_{2K}$-MPh were prepared to a weight percent concentration of 2 g L$^{-1}$ in HNO$_3$ (pH 1.5). All nanoparticle dispersions and polymer solutions were filtered with Millipore filter 0.22 μm. The dispersions were added dropwise to the polymer solutions under magnetic stirring keeping the mixing volume ratio at $X_C/5$. After increasing their pHs to 8 by addition of NH$_4$OH, the dispersions were centrifuged at 4000 rpm inside Merck centrifuge filters (pore 100000 g mol$^{-1}$) and concentrated to 20 g L$^{-1}$.

**Nanoparticle stability:** A volume of 100 μL of a 20 g L$^{-1}$ dispersion of polymer coated nanoparticles were poured and homogenized rapidly in 900 μL of phosphate buffer saline (PBS) or cell culture medium (Dulbecco's Modified Eagle's Medium, DMEM) containing 10 vol. % fetal bovine serum. Scattered intensity and diameter were measured by light scattering. After mixing, the measurements were monitored at day 1, 2, 7 and 60. Nanoparticles are considered to be stable if their hydrodynamic diameter in a given solvent remains constant as a function of the time and equal to its initial value.

**Conflicts of interest**
There are no conflicts of interest to declare.

# Acknowledgements

We thank Virginie Berthat, Geoffroy Goujon, Isabelle Margaill, Nathalie Mignet, Evdokia Oikonomou, Chloé Puisney, Caroline Roques for fruitful discussions. Alexandre Chevillot is acknowledged for his support with the thermogravimetric analysis experiments. Serge Stoll from the University Geneva is grateful acknowledged for providing us the titanium dioxide nanoparticles. ANR (Agence Nationale de la Recherche) and CGI (Commissariat à l'Investissement d'Avenir) are acknowledged for their financial support of this work through Labex SEAM (Science and Engineering for Advanced Materials and devices) ANR 11 LABX 086, ANR 11 IDEX 05 02. We acknowledge the ImagoSeine facility (Jacques Monod Institute, Paris, France), and the France BioImaging infrastructure supported by the French National Research Agency (ANR-10-INSB-04, « Investments for the future »). This research was supported in part by the Agence Nationale de la Recherche under the contract ANR-13-BS08-0015 (PANORAMA), ANR-12-CHEX-0011 (PULMONANO) and ANR-15-CE18-0024-01 (ICONS, Innovative polymer coated cerium oxide for stroke treatment).

# TOC Image

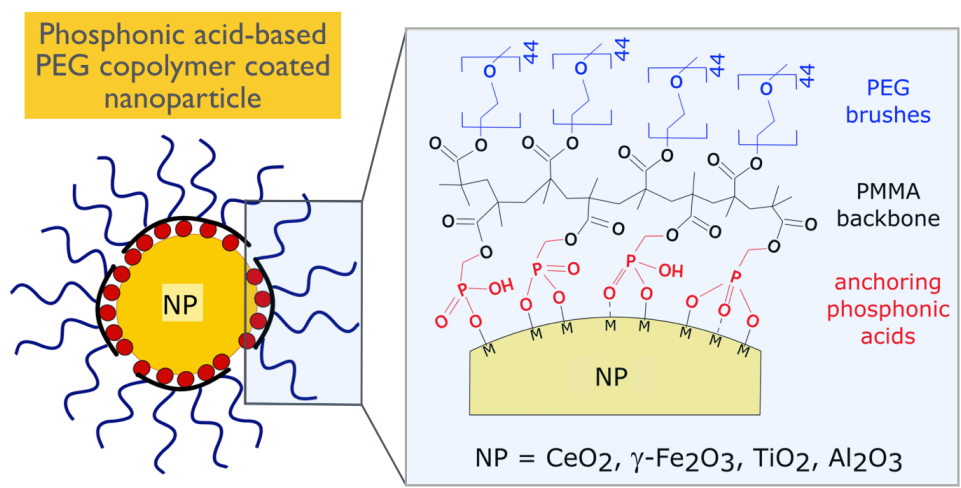